\documentclass[twoside,fleqn]{article}
\usepackage{espcrc2}
\usepackage{graphicx}
\usepackage{subeqn}

\def\openone{\leavevmode\hbox{\small1\kern-3.8pt\normalsize1}}%
\newcommand{\1}{\openone}

\def\slash#1{\setbox0=\hbox{$#1$}#1\hskip-\wd0\dimen0=5pt\advance
       \dimen0 by-\ht0\advance\dimen0 by\dp0\lower0.5\dimen0\hbox
         to\wd0{\hss\sl/\/\hss}}

\newlength{\miniwidth}
\newlength{\miniwidthplot}
\newlength{\minicolumn}
\newlength{\nseparation}
\newenvironment{nfigure}
	{\begin{figure}[htbp]\hrule\vspace{\nseparation}\par}
	{\vspace{\nseparation}\par \hrule \end{figure}}
\newenvironment{ntable}
	{\begin{table}[htbp]\hrule\vspace{\nseparation}\par}
	{\vspace{\nseparation}\par \hrule \end{table}}
\newcommand{\ncaption}[1]{\caption{\slshape #1}}

\newcommand{\lt}{\left}
\newcommand{\rt}{\right}
\newcommand{\ov}{\overline}

\newcommand{\no}{\nonumber}
\newcommand{\diff}[1]{\frac{d}{d #1}}

\newcommand{\g}{\gamma}

\newcommand{\bare}{\mathrm{bare}}
\newcommand{\ba}{\bare}
\newcommand{\eps}{\varepsilon}
\newcommand{\Lagr}{{\mathcal{L}}}
\newcommand{\lag}{\Lagr}
\newcommand{\eff}{{\mathrm{eff}}}
\newcommand{\Leff}{\Lagr_\eff^\dstwo}

\newcommand{\eq}[1]{(\ref{#1})}
\newcommand{\fig}[1]{Fig.~\ref{#1}}
\newcommand{\tab}[1]{Table~\ref{#1}}

\newcommand{\hc}{\mathrm{h.c.}}
\newcommand{\mc}{m_c^2}

\newcommand{\ms}{m_s^2}

\newcommand{\mw}{M_W^2}
\newcommand{\muc}{\mu_c}
\newcommand{\mub}{\mu_b}
\newcommand{\mut}{\mu_{tW}}
\newcommand{\tw}{\widetilde{\openone}} 
\newcommand{\topenone}{\widetilde{\openone}}
\newcommand{\as}{\alpha_s}

\newcommand{\gev}{\,\mathrm{GeV}}

\newcommand{\laMSb}{\ensuremath{\Lambda_{\overline{\mathrm{MS}}}}}
\newcommand{\laQCD}{\ensuremath{\Lambda_{\mathrm{QCD}}}}
\newcommand{\gf}{\,G_F}
\newcommand{\gft}{\, G_F^2}

\newcommand{\dstwo}{\ensuremath{\mathrm{|\Delta S| \!=\!2}}}
\newcommand{\dsone}{\ensuremath{\mathrm{|\Delta S| \!=\!1}}}

\newcommand{\msb}{\ensuremath{\ov{\textrm{MS}}}}
\newcommand{\dmu}{\mu \diff{\mu}}

\newcommand{\wt}[1]{\widetilde{#1}}

\newcommand{\stwo}{\mathrm{S2}}
\newcommand{\cll}[1]{\wt{C}_{\stwo}^{(#1)}}
\newcommand{\oll}{\ensuremath{\tilde{Q}_{\stwo}}}
\newcommand{\zll}{\wt{Z}_{\stwo}}
\newcommand{\F}{F}

\newcommand{\mr}[1]{\mathrm{#1}}

\newcommand{\leo}{\lag_\mr{eff}^\mr{|\Delta S| =  2}}
\newcommand{\oloc}{\wt{Q}_7}
\newcommand{\cloc}{\wt{C}_7}
\newcommand{\zloc}{\wt{Z}_{77}}
\newcommand{\zbi}[1]{\wt{Z}_{#1,7}}
\newcommand{\gloc}{\wt{\gamma}_{77}}
\newcommand{\gbi}[1]{\wt{\gamma}_{#1,7}}

\newcommand{\errorpm}[2]{
\raisebox{-0.5ex}{\shortstack[l]{$\scriptstyle+#1$\\$\scriptstyle-#2$}}
}
\newcommand{\pr}{Phys.\ Rev.\ }
\newcommand{\prd}{\pr D}
\newcommand{\np}{Nucl.\ Phys.\ }
\newcommand{\npb}{\np B}

\newcommand{\pl}{Phys.\ Lett.\ }

\newcommand{\prp}{preprint }

\newcommand{\hn}{S.~Herrlich and U.~Nierste}

\title{
	\textsf{
	\hfill
	\begin{minipage}{3in}
	\begin{flushright}\small 
	June 1996\\
	DESY 96-100\\
	hep-ph/9606242
	\end{flushright}
	\end{minipage}
	}\\[1.0ex]
	The Coefficient $\eta_3$ of the \dstwo -Hamiltonian in the
	Next-To-Leading Order}

\author{Stefan Herrlich
	\address{DESY-IfH, Platanenallee 6, D-15738 Zeuthen, Germany}
	\thanks{e-mail:
	\texttt{herrl@feynman.t30.physik.tu-muenchen.de} } }

\begin{document}

\begin{abstract}
I present the calculation of the QCD short distance coefficient
$\eta_3$ of the \dstwo-hamiltonian in the next-to-leading order (NLO)
of renormalization group improved perturbation theory.  It involves
the two-loop mixing of bilocal structures composed of two \dsone\
operators into \dstwo\ operators.  The next-to-leading order
corrections enhance $\eta_3$ by 27\% to
\begin{displaymath}
\eta_3=0.47\errorpm{0.03}{0.04}
\end{displaymath}
thereby affecting the phenomenology of the CP-parameter $\epsilon_K$
sizeably.  $\eta_3$ depends on the physical input parameters $m_t$,
$m_c$ and $\laMSb$ only weakly.  The quoted error stems from
factorization scale dependences, which have reduced compared to the
old leading log result.  We further discuss some field theoretical aspects
of the calculation such as the renormalization group equation for
Green's functions with two operator insertions and the renormalization
scheme dependence caused by the presence of evanescent operators.
This article is based on work done in collaboration with U.~Nierste.
\end{abstract}

\maketitle

\setcounter{footnote}{0}

\section{Introduction}\label{sect:intro}

The effective low-energy hamiltonian inducing the \dstwo-transition
reads:
\begin{eqnarray}
H^{\dstwo} &=&
	\lambda_c^2 H^c + \lambda_t^2 H^t + 2\lambda_c\lambda_t H^{ct}
\no \\
&=&
	\frac{\gft}{16\pi^2} \mw \biggl[
	\lambda_c^2 \eta_1^\star
		S\!\left(x_c^\star\right)
	+
	\lambda_t^2 \eta_2^\star
		S\!\left(x_t^\star\right)
\no \\
&&\hspace{1em}
	+2\lambda_c \lambda_t \eta_3^\star
		S\!\left(x_c^\star,x_t^\star\right)
	\biggr]
	b\!\left(\mu\right) \oll\!\left(\mu\right)
\no \\
&&
	+ \hc
\label{hamiltonian}
\end{eqnarray}
Here $\gf$ denotes Fermi's constant, $M_W$ is the W boson mass,
$\lambda_j = V_{jd} V_{js}^{*}, j=c,t$ comprises the CKM-factors, and
$\oll$ is the local dimension-six \dstwo\ four-quark operator
\begin{equation}
\oll =
	\left[\ov{s} \gamma_\mu \left(1-\gamma_5\right) d\right]
	\cdot
	\left[\ov{s} \gamma_\mu \left(1-\gamma_5\right) d\right]
\label{oll}
\end{equation}
The $x_q^\star = {m_q^\star}^2/\mw$, $q=c,t$ encode the running quark
masses $m_q^\star=m_q(m_q)$ in the \msb\ scheme.  In writing
{\eq{hamiltonian}} the GIM mechanism $\lambda_u+\lambda_c+\lambda_t=0$
has been used to eliminate $\lambda_u$.  Further we have set $m_u=0$.
The Inami-Lim functions $S(x)$, $S(x,y)$ describe the
{\dstwo}-transition amplitude in the absence of QCD.  They read:
\begin{subequations}
\label{sfkts}
\begin{eqnarray}
S(x_t) &=& x_t \lt[ \frac{1}{4} + \frac{9}{4}
\frac{1}{1-x_t} - \frac{3}{2} \frac{1}{(1-x_t)^2} \rt]
\no \\
&& - \frac{3}{2} \lt[ \frac{x_t}{1-x_t} \rt]^3 \ln x_t,
\label{sxt} \\
S(x_c) &=& x_c + O(x_c^2),
\label{ilgim} \\
S(x_c,x_t) &=& - x_c \ln x_c + x_c \F(x_t)
\no \\
&& + O(x_c^2 \ln x_c),
\label{sxcxt}
\end{eqnarray}
\end{subequations}
with
\begin{eqnarray}
\F(x_t)&=&  \frac{x_t^2-8 x_t+4 }{4 (1-x_t)^2} \ln x_t
            + \frac{3}{4} \frac{x_t}{x_t-1}.
\label{deff}
\end{eqnarray}
In \eq{ilgim} and \eq{sxcxt} we have only kept terms which are larger
than those of order $(m_s m_c)/\mw$ neglected by setting the external
momenta to zero.

In \eq{hamiltonian} the short-distance QCD corrections are comprised
in the coefficients $\eta_1^\star$, $\eta_2^\star$ and $\eta_3^\star$
with their explicit dependence on the renormalization scale $\mu$
factored out in the function $b(\mu)$.  In absence of QCD corrections
$\eta_i^\star b(\mu) = 1$.

{\tab{tab:rglogs}} summarizes the logarithms summed by the forthcoming
renormalization group (RG) evolution from $M_W$ down to $m_c$ in the
different orders.
\begin{ntable}
\ncaption{Logarithms summed by the RG evolution from $M_W$ down to
$m_c$ for the three terms in \eq{hamiltonian}, $n=0,1,2,\ldots$ The
last line shows the order in which the dependence on $m_t$ enters.}
\label{tab:rglogs}
\begin{tabular}{@{}l@{}*{3}{@{\hspace{0.5em}}r@{}c@{}l@{}}}
Order &
& $H^c$ &&& $H^t$ &&& $H^{ct}$ & \\
\hline
LO&
 &$(\as\ln\!x_c)^n$&&
 &$(\as\ln\!x_c)^n$&&
 &$(\as\ln\!x_c)^n$&$\ln\!x_c$
\\
NLO&
$\as$&$(\as\ln\!x_c)^n$&&
$\as$&$(\as\ln\!x_c)^n$&&
&$(\as\ln\!x_c)^n$&
\\ \hline
$m_t$ &
 & none &&
 & in LO &&
 & in NLO &
\end{tabular}
\end{ntable}

The first complete determination of the coefficients $\eta_i$,
$i=1,2,3$ in the leading order (LO) is due to Gilman and Wise
{\cite{gw}}.  However, the LO expressions have several conceptual
drawbacks:
\begin{enumerate}
\item The fundamental QCD scale parameter \laMSb\ is not
well-defined in the LO.
\item The quark mass dependence of the $\eta_i$'s is not
correctly reproduced by the LO expressions.  Especially the
$m_t^\star$-dependent terms in $\eta_3^\star \cdot
S(x_c^\star,x_t^\star)$ already belong to the NLO.
\item Similarly the question of the \emph{definition} of the quark
masses (i.e.\ the renormalization scheme and scale) to be used in
{\eq{hamiltonian}} is a next-to-leading order issue.
\item The LO results for $\eta_1$ and $\eta_3$ show a large dependence
on the factorization scales, at which one integrates out heavy
particles.  In the NLO these uncertainties are reduced considerably.
\item One must go to the NLO to judge whether perturbation theory 
works, i.e.\ whether the radiative corrections are small.  After all
the corrections can be sizeable.
\end{enumerate}
To overcome the limitations listed above one has to go to the
next-to-leading order (NLO).  This programme has been started with the
calculation of $\eta_2^\star$ by Buras, Jamin and Weisz \cite{bjw}.
Then Nierste and I have derived the NLO expressions for $\eta_1^\star$
{\cite{hn1}} and $\eta_3^\star$ \cite{hn}.  This article will
essentially deal with the term $H^{ct}$ of \eq{hamiltonian} and is
based on \cite{hn}.

\section{The NLO calculation of $\eta_3$ above the charm threshold}
\label{sect:above}

Here we will shortly describe how the large logarithm $\ln x_c$
present in \eq{sxcxt} is summed to all orders in perturbation theory.
This is done in two steps: First one sets up an effective lagrangian
$\Leff$ in which the W boson and the top quark are removed as dynamic
degrees of freedom.  In $\Leff$ the \dsone\ and \dstwo\ transitions
are described by local four-quark operators, which are multiplied by
Wilson coefficients.  The general structure of $\Leff$ reads:
\begin{eqnarray}
\Leff &=&
-\frac{\gf}{\sqrt{2}} V_{\mathrm{CKM}} \sum_{k} C_{k} Q_k
\no \\
&&
-\frac{\gf^2}{2} V_{\mathrm{CKM}} \sum_{l} \wt{C}_{l} \wt{Q}_{l}.
\label{lgeneric}
\end{eqnarray}
Here the $V_{\mathrm{CKM}}$ denote products of CKM elements.  The
$Q_k$, $\wt{Q}_l$ represent local \dsone\ and \dstwo\ operators and
the $C_k$, $\wt{C}_l$ are the corresponding Wilson coefficient
functions.  The \dsone\ part of \eq{lgeneric} contributes to \dstwo\
transitions via diagrams of the type displayed in \fig{fig:cc-cc-lo}.
\begin{nfigure}
\includegraphics[clip,width=\minicolumn]{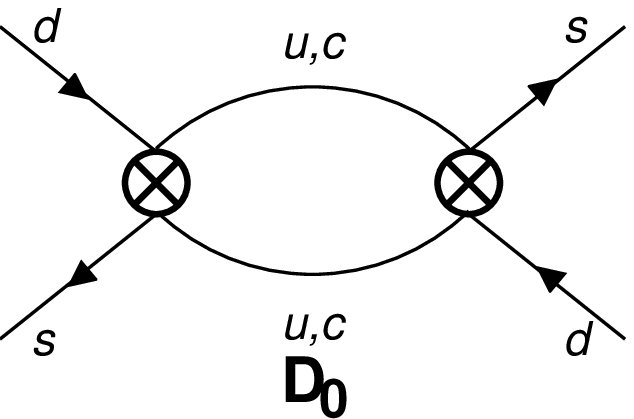}
\hfill
\includegraphics[clip,width=\minicolumn]{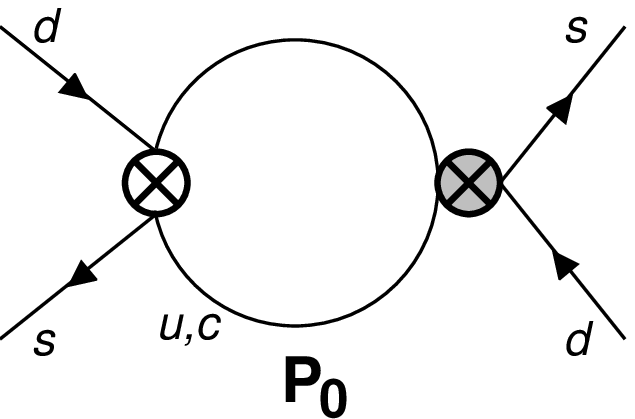}
\ncaption{
The diagrams $\mathsf{D_0}$ and $\mathsf{P_0}$ in the effective
five- and four-quark theory.  The light crosses denote insertions of
local {\dsone}\ current-current operators, the grey ones insertion of
local {\dsone}\ four-quark penguin operators.}
\label{fig:cc-cc-lo}
\end{nfigure}
The Wilson coefficients are fixed at a factorization scale
$\mut=O(M_W,m_t)$ by the requirement that the \dsone\ and \dstwo\
Green's functions derived from \eq{lgeneric} are equal to the same
quantities calculated with the full SM lagrangian.  The effect of this
procedure is that the $\ln x_c$ present in \eq{sxcxt} is split as
$\ln(\mut/M_W)+\ln(m_c/\mut)$.  Here the former term resides in the
Wilson coefficients while the latter is contained in the matrix
elements of the local operators.  The choice $\mut=O(M_W,m_t)$ ensures
that the Wilson coefficients do not contain large logarithms and
therefore can be reliably calculated in ordinary perturbation theory.

The second step is the RG evolution of the Wilson coefficients from
the scale $\mut$ down to $\muc=O(m_c)$ which sums
$\ln(\muc/\mut)$.\footnote{For simplicity we ignore the intermediate
scale $\mub=O(m_b)$ at which the bottom quark gets integrated out.}

\subsection{The Operator Basis}\label{sect:ops}

Let us now construct $\Leff$ in \eq{lgeneric} as far it is needed for
the calculation of $\eta_3$.  Since the presence of \dstwo\ terms in
\eq{lgeneric} does not affect the \dsone\ part of $\Leff$, we can
simply take the latter from \cite{bw,bjlw,bjlw2}.  They consist of the
following set of operators:
\begin{subequations}
\label{SOneOps}
\begin{eqnarray}
Q_1^{kl} &=&
\left(\bar{s} \gamma_{\mu} L k\right) \cdot
\left(\bar{l} \gamma^{\mu} L d\right) \cdot
\topenone
\label{defQ1} ,
\\
Q_2^{kl} &=&
\left(\bar{s} \gamma_{\mu} L k\right) \cdot
\left(\bar{l} \gamma^{\mu} L d\right) \cdot
\openone
\label{defQ2} ,
\\
Q_3 &=& \left(\ov{s} \g_\mu L d\right) \cdot \sum_{q=d,u,s,\ldots}
\left(\ov{q}\g^\mu L q\right) \cdot \1
\label{defQ3} ,
\\
Q_4 &=& \left(\ov{s} \g_\mu L d\right) \cdot \sum_{q=d,u,s,\ldots}
\left(\ov{q}\g^\mu L q\right) \cdot \tw
\label{defQ4} ,
\\
Q_5 &=& \left(\ov{s} \g_\mu L d\right) \cdot \sum_{q=d,u,s,\ldots}
\left(\ov{q}\g^\mu R q\right) \cdot \1
\label{defQ5} ,
\\
Q_6 &=& \left(\ov{s} \g_\mu L d\right) \cdot \sum_{q=d,u,s,\ldots}
\left(\ov{q}\g^\mu R q\right) \cdot \tw
\label{defQ6} .
\end{eqnarray}
\end{subequations}
Here the $Q_i^{kl}$, $i\!=\!1,2$, $k,l\!=\!u,c$ represent the \dsone\
current-current operators, the $Q_i$, $i\!=\!3,\ldots,6$ the
QCD-penguin operators.  The sum in \eq{defQ3}--\eq{defQ6} runs over
all active flavours.  Further $L,R = (1\mp\gamma_5)$ and $\openone$,
$\topenone$ denote color singlet and anti-singlet, i.e.\
$Q_1^{kl}=\left(\ov{s}_i \gamma_\mu L k_j\right) \cdot \left(\ov{l}_j
{\gamma^\mu} L d_i\right)$ with $i,j$ being color indices.

Further we need the \dstwo\ operators present in \eq{lgeneric}.
Consider first the diagram $\mathsf{D_0}$ of \fig{fig:cc-cc-lo} with
two internal charm quarks and zero external momenta.  The only
physical dimension-eight \dstwo\ operator required to absorb its
divergence reads
\begin{eqnarray}
\oloc &=& \frac{\mc}{g^2 \mu^{2 \eps}} \oll
= \frac{\mc}{g^2 \mu^{2 \eps}} \cdot
     \ov{s} \g_\mu L d \cdot \ov{s} \g^\mu L d ,
\label{defQ7}
\end{eqnarray}
which follows from power counting and the absence of any non-zero mass
parameter apart from $m_c$.  The inverse powers of $g$ are introduced
for later convenience as in {\cite{gw}}. One may arbitrarily shift
such factors from the Wilson coefficient into the definition of the
operator. The factor $\mu^{- 2\eps}$ stems from $g_{\mr{bare}}= Z_g g
\mu^{\eps}$ and the fact that
\begin{eqnarray}
\oloc^\ba &=& \frac{ m_{c,\,\ba }^{2} }{g_{\ba }^{ 2}} \lt[
     \ov{s} \g_\mu L d \cdot \ov{s} \g^\mu L d \rt]^{\ba} .
\label{q7bare}
\end{eqnarray}
must be independent of $\mu$.\footnote{Here and in the following the
superscript ``bare'' denotes unrenormalized operators, while
renormalized ones do not carry an additional superscript.}  Any other
dimension-eight \dstwo\ operator contains one or two powers of $m_c$
less than $\oloc$ and derivatives and/or gluon fields instead.  Their
on-shell matrix elements are suppressed by powers of $m_s/m_c$ with
respect to those of $\oloc$, so that they do not contribute to the
coefficient of the leading dimension-six operator \emph{below} the
charm threshold (cf.\ (\ref{hamiltonian})).  Likewise they cannot mix
with $\oloc$ under renormalization.

Therefore our operator basis consists of $Q_{1,2}^{kl}$,
$Q_{3,\ldots,6}$ and $\oloc$\footnote{In addition to these physical
operators we have to take into account several evanescent operators.
This class of operators appears quite naturally when one has to deal
with the renormalization of operators containing more than one fermion
line in dimensional regularization {\cite{bw,dg,hn2}}.  To illustrate
some of the findings of \cite{hn2} we have defined the evanescent
operators with some arbitrary coefficients $a_1$, $a_2$, $\wt{a}_1$,
$\wt{b}_1$:
\begin{subequations}
\label{StdEvas}
\begin{eqnarray}
E_1[Q_j]\hspace{-0.7em} &\! = \! & \bigl[
\gamma_\mu \gamma_\nu \gamma_\eta L \otimes \gamma^\eta \gamma^\nu \gamma^\mu L
\no \\ &&
- \left(4+a_1\eps\right) \gamma_\mu L \otimes \gamma^\mu L
\bigr] K_{1j} , \; j=1,\ldots 4,
\label{StdEvas1-12}\\
E_1[Q_j]\hspace{-0.7em} &\! = \! & \bigl[
\gamma_\mu \gamma_\nu \gamma_\eta R \otimes \gamma^\eta \gamma^\nu \gamma^\mu L
\no \\ &&
- \left(16+a_2\eps\right) \gamma_\mu R \otimes \gamma^\mu L
\bigr] K_{1j} , \; j=5,6 ,
\label{StdEvas1-56}\\
E_1[\oloc]\hspace{-0.7em} &\! = \! & \frac{\mc}{g^2} \bigl[
\gamma_\mu \gamma_\nu \gamma_\eta L \otimes \gamma^\eta \gamma^\nu \gamma^\mu L
\no \\ &&
- \left(4+\tilde{a}_1\eps\right) \gamma_\mu L \otimes \gamma^\mu L
\bigr] K_{12} ,
\label{StdEvas1-loc}\\
E_2[\oloc]\hspace{-0.7em} &\! = \! & \frac{\mc}{g^2} \bigl[
\gamma_\mu \gamma_\nu \gamma_\eta \gamma_\sigma \gamma_\tau L \otimes
\gamma^\tau \gamma^\sigma \gamma^\eta \gamma^\nu \gamma^\mu L
\no \\ &&
- \left[\left(4+\tilde{a}_1\eps\right)^2+\tilde{b}_1 \eps \right]
\gamma_\mu L \otimes \gamma^\mu L \bigr] K_{22}.
\label{StdEvas2-loc}
\end{eqnarray}
\end{subequations}
with the color factors
\begin{equation}
\begin{array}{@{}l*{3}{@{\;=\;}l}}
K_{12}&K_{13}&K_{15}&\frac{1}{2} \topenone - \frac{1}{2 N} \openone,\\
K_{11}&K_{14}&K_{16}&\frac{1}{4} \openone + \frac{N^2-2}{4 N}\topenone.
\end{array}
\label{DefColourEva}
\end{equation}
Apart from places where it is indicated we will always state the
results corresponding to
\begin{eqnarray}
a_1 = -8, \hspace{3em}
a_2 = -16, \hspace{3em}
\tilde{a}_1 = -8,
\label{StdEvasNum}
\end{eqnarray}
in order to comply with the standard choice used in
\cite{bjw,hn1,bw,bjlw,bjlw2}.  Since NLO anomalous dimensions
and matching corrections of physical operators do not depend
on $\tilde{b}_1$, we do not give a numerical value.  Likewise
we do not need the value of the colour factor $K_{22}$.
}
and $\Leff$ is found as:
\begin{eqnarray}
\Leff\hspace{-1em}&\!\!\!\!=\!\!\!\!&
- \frac{\gf}{\sqrt{2}} \sum_{i=1}^6  C_i
\biggl[  \sum_{j=1}^2\hspace{-0.3em} Z^{-1}_{ij}\hspace{-0.5em}
\sum_{k,l =u,c}\hspace{-0.5em} V_{ks}^\ast  V_{ld}   Q_j^{kl,\bare}
\no \\
&& \hspace{6em}
- \lambda_t \sum_{j=3}^6 Z^{-1}_{ij} Q_j^{\bare}  \biggr] \no \\
&& - \frac{\gft}{2} \lambda_c \lambda_t
    \biggl[ \sum_{k=1}^{2} \sum_{l=1}^{6}
   C_k C_l \zbi{kl}^{-1}
\no \\
&& \hspace{6em}
+ \cloc \zloc^{-1} \biggr]
   \oloc^{\bare}
   \nonumber \\
&& + \parbox[t]{14em}{counterterms proportional to unphysical operators.}
\label{lags2}
\end{eqnarray}

\subsection{The Effective Lagrangian $\Leff$ at the Scale $\mut$}
\label{sect:mutW}

Besides the operators we need to know their corresponding Wilson
coefficient functions at the initial scale $\mut$.  The ones for the
\dsone\ case, $C_i$, $i\!=\!1,\ldots,6$ can be taken from
\cite{bw,bjlw,bjlw2}.  It is important to note that only $C_2$ starts
at $O(\as^0)$, the others at $O(\as^1)$.  This means that the NLO
matching can be done solely with the diagram $\mathsf{D_0}$ of
{\fig{fig:cc-cc-lo}} with two insertions of $Q_2$.  One easily finds
\begin{equation}
\cloc(\mu_{tW}) = \left\{
\begin{array}{@{}ll}
0 & \mbox{in LO} \\
\displaystyle \frac{\as(\mu_{tW})}{4\pi} \bigl(
-8\ln\frac{\mu_{tW}}{M_W} & \\
\hspace{0.5em}+4F\left(x_t\left(\mu_{tW}\right)\bigr)
+ 2
\right) & \mbox{in NLO}
\end{array}
\right.\hspace{-1.5em},
\label{clocInit}
\end{equation}
where $F(x_t)$ is the top dependent part of $S(x_c,x_t)$ defined in
{\eq{deff}}.  The factor $\as$ originates from the special definition
of $\oloc$ in \eq{defQ7}.  Note how the large logarithm $\ln x_c$ in
\eq{sxcxt} is split between the Wilson coefficient $\cloc$ and the
matrix element.  The NLO result in {\eq{clocInit}} is specific to the
NDR scheme with {\eq{StdEvasNum}}.\footnote{
The Wilson coefficient $\cloc$ depends on $\wt{a}_1$:
\begin{displaymath}
\cloc(\mu_{tW}) =
\frac{\as(\mu_{tW})}{4\pi} \left[
-8\ln\frac{\mu_{tW}}{M_W}+4F\left(x_t\left(\mu_{tW}\right)\right)
-\left(6+\tilde{a}_1\right)
\right]
\end{displaymath}
}

\subsection{Evolving down $\Leff$ from $\mut$ to $\muc$}\label{sect:evol}

Next we have to evolve down the Wilson coefficients present in $\Leff$
{\eq{lags2}} from $\mut$ to $\muc$.  For the \dsone\ functions $C_i$,
$i\!=\!1,\ldots,6$ this is achieved by standard methods \cite{bjlw}.
To calculate the running of the \dstwo\ coefficient $\cloc$ we need to
derive and solve the corresponding RG equation.  From $\dmu\Leff=0$ in
{\eq{lags2}} one finds
\begin{eqnarray}
\dmu \cloc\left(\mu\right) \hspace{-1em} &=& \hspace{-1em}
\cloc\left(\mu\right) \gloc
\no \\
&&\hspace{-1em}+ \sum_{k=1}^2 \sum_{k^\prime =1}^6 
   C_{k}\left(\mu\right) C_{k'}\left(\mu\right) \gbi{kk'}
\label{RGdouble}
\end{eqnarray}
with the \emph{anomalous dimension tensor}\footnote{``Anomalous
dimension tensor'' is clearly a misnomer and only used to distinguish
$\gbi{kn}$ from ordinary anomalous dimension (square) matrices.}
\begin{eqnarray}
\gbi{kn}\hspace{-1em} &=& \hspace{-0.8em}
\frac{\as}{4\pi} \gbi{kn}^{\left(0\right)}
+ \left(\frac{\as}{4\pi}\right)^2 \gbi{kn}^{\left(1\right)}
+ \ldots
\no \\
&=&\hspace{-0.8em}
- \sum_{k^\prime=1}^2 \sum_{n^\prime =1}^6 
\left[\gamma_{kk'} \delta_{nn'} + \delta_{kk'} \gamma_{nn'}\right]
\zbi{k'n'}^{-1} \zloc
\no \\
&&-\left[\dmu \zbi{kn}^{-1}\right] \zloc .
\label{AnomDimTens}
\end{eqnarray}
It is possible to solve the inhomogeneous equation \eq{RGdouble}
directly but this turns out to be inconvenient for practical purposes.
Instead we may combine the evolution equations of the \dsone\ and
{\dstwo}\ Wilson coefficients into a single matrix equation.  This is
made possible because the GIM mechanism ensures that at least one of
the two \dsone\ operator insertions in diagrams like the ones
displayed in \fig{fig:cc-cc-lo} is of the current-current type.  The
current-current part of the {\dsone}\ mixing matrix, i.e.\ the entries
related to $Q_1$, $Q_2$, can be diagonalized exactly using the basis
\begin{equation}
Q_\pm^{kl} = \frac{1}{2} \left(Q_2^{kl} \pm Q_1^{kl}\right).
\label{DefQpm}
\end{equation}
Then \eq{RGdouble} together with the RG equation of the \dsone\ Wilson
coefficients splits into two independent inhomogeneous RG equations
\begin{equation}
\dmu \cloc^{\pm}\left(\mu\right) =
	\gloc \cloc^{\pm}\left(\mu\right)
	+ \gbi{\pm k}  C_\pm\left(\mu\right) C_k\left(\mu\right).
\label{RGdoubleSep}
\end{equation}
Here the decomposition of $\cloc(\mu_{tW})$ into $\cloc^{\pm}(\mu_{tW})$
is completely arbitrary provided one satisfies
\begin{equation}
\cloc\left(\mu_{tW}\right) =
\cloc^{+}\left(\mu_{tW}\right) + \cloc^{-}\left(\mu_{tW}\right).
\label{clocdec}
\end{equation}
This decomposition is then automatically  preserved at any
renormalization scale.

Each of the two equations in \eq{RGdoubleSep} may be written as a
7$\times$7 matrix equation, which may be solved by standard methods.
We can even do better and collapse the two resulting 7$\times$7 
matrix equations into one 8$\times$8 matrix equation:
\begin{equation}
\dmu \vec{D} =
\widehat{\gamma}^T \cdot \vec{D}
\label{RGdouble1matrix8}
\end{equation}
with
\begin{subequations}
\label{RGdouble8}
\begin{eqnarray}
\widehat{\gamma}^T \hspace{-0.8em}&=&\hspace{-1.0em}
\left(
\begin{array}{ccc}
\gamma^T & 0 & 0 \\
\gbi{+}^{T} & \gloc - \gamma_{+} & 0 \\
\gbi{-}^{T} & 0 & \gloc - \gamma_{-}
\end{array}
\right)
\label{RGdouble8Anom}
\\
\gbi{\pm}^{T} \hspace{-0.8em}&=&\hspace{-1.0em}
\left(
\gbi{\pm1\!}, \gbi{\pm2\!}, \gbi{\pm3\!},
\gbi{\pm4\!}, \gbi{\pm5\!}, \gbi{\pm6\!}
\right)
\label{RGdouble8Tens}
\\
\vec{D}\left(\mu\right) \hspace{-0.8em}&=&\hspace{-1.0em}
\left(
\begin{array}{c}
\vec{C}\left(\mu\right) \\
\cloc^{+}\left(\mu\right) / C_{+}\left(\mu\right) \\
\cloc^{-}\left(\mu\right) / C_{-}\left(\mu\right)
\end{array}
\right).
\label{RGdouble8WC}
\end{eqnarray}
\end{subequations}

We now need to know the elements of the anomalous dimension tensor
$\gbi{\pm i}$, $i=1,\ldots,6$.  They are obtained from the
renormalization constants using the definition in \eq{AnomDimTens} and
expanding the quantities in there in powers of $\as$ and $1/\eps$.
Here it is important to include the finite renormalization terms
needed for the correct treatment of the evanescent operators
{\eq{StdEvas}}.  To calculate the LO term $\gbi{\pm i}^{(0)}$ one
needs to know the $1/\eps$ parts of the one-loop diagrams displayed in
{\fig{fig:cc-cc-lo}}, the NLO part $\gbi{\pm i}^{(1)}$ requires the
evaluation of a set of two-loop graphs.  We find:
\begin{subequations}
\label{ResAnomTens}
\begin{eqnarray}
\gbi{+}^{(0)} =
\left(
\begin{array}{r}
-16 \\ -8 \\ -32 \\ -16 \\ 32 \\ 16
\end{array}
\right),
&&\hspace{-1em}
\gbi{-}^{(0)} =
\left(
\begin{array}{r}
8 \\ 0 \\ 16 \\ 0 \\ -16 \\ 0
\end{array}
\right),
\label{ResAnomTens0}
\\
\gbi{+}^{(1)} =
\left(
\begin{array}{r}
-212 \\ -28 \\ -456 \\ -88 \\ \frac{1064}{3} \\ \frac{832}{3}
\end{array}
\right),
&&\hspace{-1em}
\gbi{-}^{(1)} =
\left(
\begin{array}{r}
276 \\ -92 \\ 520 \\ -216 \\ -\frac{1288}{3} \\ 0
\end{array}
\right).
\label{ResAnomTens1}
\end{eqnarray}
\end{subequations}
As usual the NLO anomalous dimension tensor depends on the
renormalization scheme.  The result {\eq{ResAnomTens1}} corresponds to
the NDR scheme with the definition of the evanescent operators
corresponding to {\eq{StdEvasNum}}.\footnote{In our two-loop
calculation we have kept $a_1,a_2, \wt{a}_1$ and $\wt{b}_1$ in
\eq{StdEvas} arbitrary yielding
\begin{subequations}
\begin{eqnarray}
\gbi{+}^{(1)} &=&
\left(
\begin{array}{@{}*{4}{r@{}}}
-\frac{188}{3} & -\frac{74}{3} a_1 & & + \frac{130}{3} \tilde{a}_1 \\
-\frac{100}{3} & -\frac{34}{3} a_1 & & + \frac{32}{3} \tilde{a}_1 \\
-\frac{1816}{3} & - \frac{88}{3} a_1 & & + \frac{32}{3} \tilde{a}_1 \\
-\frac{680}{3} & - \frac{80}{3} a_1 & & + \frac{28}{3} \tilde{a}_1 \\
\frac{1576}{3} & + \frac{80}{3} a_1 & + \frac{8}{3} a_2 &
	- \frac{32}{3} \tilde{a}_1 \\
\frac{1664}{3} & + \frac{28}{3} a_1 & + \frac{52}{3} a_2 &
	- \frac{28}{3} \tilde{a}_1
\end{array}
\right),
\\
\gbi{-}^{(1)} &=&
\left(
\begin{array}{@{}*{4}{r@{}}}
\frac{124}{3} &+\frac{22}{3} a_1 & & - \frac{110}{3} \tilde{a}_1 \\
-12 &+6 a_1 & & + 4 \tilde{a}_1 \\
\frac{1496}{3} & + \frac{8}{3} a_1 & & - \frac{16}{3} \tilde{a}_1 \\
-120 & + 8 a_1 & & + 4 \tilde{a}_1 \\
\frac{1160}{3} & - \frac{16}{3} a_1 & + \frac{8}{3} a_2 &
	+ \frac{16}{3} \tilde{a}_1 \\
-128 & - 4 a_1 & - 4 a_2 & - 4 \tilde{a}_1
\end{array}
\right)
\end{eqnarray}
\end{subequations}
}

\section{The NLO calculation of $\eta_3$ below the charm threshold}
\label{sect:below}

\subsection{The Effective Lagrangian $\Leff$ at the Scale
	$\muc=O(m_c)$}
\label{sect:muc}

After integrating out the charm quark all dependence on $m_c$
belongs to the Wilson coefficients.  This implies that the
term involving $\oloc$ in \eq{lags2} has to disappear from the
effective lagrangian, because $\oloc$ contains $m_c$ in its definition
{\eq{defQ7}}.  Further the \dsone\ operators are neglected in the
new effective lagrangian, because the matrix elements of double
insertions of these operators are at most proportional to $\ms$
rather than $\mc$.  We have already neglected such terms in all
preceding steps.

Therefore the new effective lagrangian to describe the physics below
$\muc$ reads:
\begin{eqnarray}
\leo &=&
-\frac{\gft}{16\pi^2} \biggl[
\lambda_c^2 \cll{c}\left(\mu\right)
+\lambda_t^2 \cll{t}\left(\mu\right)
\no \\
&&
+\lambda_c \lambda_t \cll{ct}\left(\mu\right)
\biggr] \zll^{-1}\left(\mu\right) \oll^\bare.
\label{lags2c}
\end{eqnarray}
This lagrangian already resembles $-H^{\dstwo}$ introduced in
{\eq{hamiltonian}}.  For the matching we have to set the Green's
function derived from \eq{lags2} and the one derived from \eq{lags2c}
equal at the scale $\mu=\muc$.

Let us start with the matching of $\cloc$ in the LO: since the
definition of $\oloc$ \eq{defQ7} contains the factor $\as$ with
respect to $\oll$ we develop an explicit inverse power of $\as$ for
the Wilson coefficient:
\begin{eqnarray}
\cll{ct}\left(\mu_c\right) &=&
\frac{\mc\left(\mu_c\right)}{2} \frac{4\pi}{\as\left(\mu_c\right)}
\cloc\left(\mu_c\right)
\mbox{in LO}.
\label{matchCct7}
\end{eqnarray}
Diagrams containing double insertions are of order $\as^0$ and
therefore start contributing to $\cll{ct}$ in the NLO.  The NLO
version of $\cll{ct}$ can be written
\begin{eqnarray}
\cll{ct}\left(\mu_c\right) &=&
\mc\left(\mu_c\right) \biggl[\frac{1}{2}
\frac{4\pi}{\as\left(\mu_c\right)} \cloc\left(\mu_c\right)
\no \\
&&\hspace{-3.5em}
+\sum_{i=+,-} \sum_{j=1}^{6} r_{ij,S2}\left(\mu_c\right)
C_i\left(\mu_c\right) C_j\left(\mu_c\right) \biggr].
\label{cllNLOct}
\end{eqnarray}

The coefficients $r_{ij,S2}\left(\mu_c\right)$ in \eq{cllNLOct} are
given by the finite parts of the diagrams in {\fig{fig:cc-cc-lo}}.  We
find:
\begin{eqnarray}
r_{ij,S2}\left(\mu_c\right)
\hspace{-1em}&=&\hspace{-1.2em}
\left\{\hspace{-0.7em}
\begin{array}{l@{\hspace{0.3em}\mbox{for }j=\hspace{-0.2em}}l}
\displaystyle
\left[-4\ln\frac{m_c(\mu_c)}{\mu_c} -1\right]
\hspace{-0.2em}\tau_{ij} & 1,2, \\[2.8ex]
\displaystyle
\left[-8\ln\frac{m_c(\mu_c)}{\mu_c} -4\right]
\hspace{-0.2em}\tau_{ij} & 3,4, \\[2.8ex]
\displaystyle
\left[8\ln\frac{m_c(\mu_c)}{\mu_c} +4\right]
\hspace{-0.2em}\tau_{ij} & 5,6,
\end{array}
\right.
\label{Rmatch}
\end{eqnarray}
where the $\tau_{ij}$'s denote the colour factors
\begin{eqnarray}
\begin{array}{l}
\displaystyle
\tau_{\pm1} = \tau_{\pm3} = \tau_{\pm5} = \frac{1 \pm N}{2}, \\
\displaystyle
\tau_{+2} = \tau_{+4} = \tau_{+6} = 1, \\
\displaystyle
\tau_{-2} = \tau_{-4} = \tau_{-6} = 0.
\end{array}
\label{DefTau}
\end{eqnarray}
Note that $r_{ij,S2}$ for $j=1,2$ depends on the definition of the
evanescent operator $E_1[\oloc]$. As usual {\eq{Rmatch}} only holds in
the NDR scheme.

\subsection{Evolving $\Leff$ below $\muc$}\label{sect:evol3}

The RG running in the effective three quark theory is particularly
simple because there is only one operator left: $\oll$.  The running
of the corresponding Wilson coefficient function $\cll{ct}$ can be
taken from \cite{bw,bjw}.

$\eta_3$ can then be determined from the identification
\begin{equation}
\Leff = - H^\dstwo
\label{identification}
\end{equation}
with $\Leff$ from \eq{lags2c} and $H^\dstwo$ from \eq{hamiltonian}.
The result of this can be found in \cite[section~5]{hn}.

\section{Numerical Results}\label{sect:num}

Let us now discuss the numerical implications of the calculation
presented in the preceding sections.  We will present the dependence
of $\eta_3^\star$ on its various physical parameters and on the
renormalization scales at which particles are integrated out.

$\eta_3^\star$ depends on the scales $\mut$, $\mub$ and $\muc$.
Further it is a function of the masses $m_t$, $m_c$ and of the QCD
scale parameter $\laQCD$.  To establish a starting point let us pick a
basic set of input parameters
\begin{eqnarray}
\begin{array}{*{2}{l}}
\begin{array}{@{}l}
m_c(m_c) = \mu_c
\\
\hspace{1em}= 1.3\gev,
\end{array}
&
\begin{array}{@{}l}
\laMSb = 0.31\gev,
\\
\laQCD^\mathrm{LO} = 0.15\gev,
\end{array}
\\
\mu_b = 4.8\gev,
& M_W = 80\gev,
\\
\mu_{tW} = 130\gev,
& m_t(m_t) = 167\gev.
\end{array}
\label{InputParam}
\end{eqnarray}
In the following $\laQCD$ is always understood to be defined with
respect to four active flavours, the corresponding quantities in
effective three- and five flavour-theories are obtained by imposing
continuity on the coupling $\as$ at $\muc$ and
$\mub$.\footnote{Threshold corrections appearing for $\mu_q\neq m_q$
are numerically negligible.}

The value for $\eta_3^\star$ corresponding to the set \eq{InputParam}
reads:
\begin{equation}
{\eta_3^\star}^\mathrm{LO} = 0.365,
\hspace{2em}
{\eta_3^\star}^\mathrm{NLO} = 0.467.
\label{Num3s}
\end{equation}
Hence the NLO calculation has enhanced $\eta_3^\star$ by 27\%.  From
the difference of 0.102 between the two values in \eq{Num3s} 0.022
originates from the change from the LO to the NLO running $\as$.  The
smallness of this contribution is caused by the adjustment of
$\laQCD^\mathrm{LO}$ to fit the NLO running coupling.  The explicit
$O(\as)$ corrections from the NLO mixing and matching contribute
0.080.

Let us further quantify the influence of the penguin operators
$Q_{3,\ldots,6}$: If one neglects them completely, one obtains
${\eta_3^\star}^\mathrm{NLO,np}=0.472$ with the set in
{\eq{InputParam}}, i.e.\ their contribution is of the order of 1\%.

\subsection{Scale Dependence of $\eta_3^\star$}\label{sect:num-scale}

Ideally $\eta_3^\star$ should not depend on the factorization scales
$\mut$, $\mub$, $\mut$.  Yet due to the truncation of the perturbation
series such a dependence shows up.  It may serve as an estimate of the
theoretical error of the calculation.

It turns out that the dependence of $\eta_3^\star$ on $\mub$ is
extremely mild.  This is due to the fact that no diagrams containing
internal bottom quarks contribute to the {\dstwo}\ process in order
$\as$.  The only places where $\mu_b$ enters are a) the running of
$\as$, b) the NLO matching matrices and anomalous dimensions of the
{\dsone}\ penguin operators.  Numerically one finds that
$\eta_3^\star$ is shifted by 0.1--0.2\% if one choses the extreme
values of $\mub \equiv \muc = O(m_c)$ or $\mub \equiv \mut =
O(m_W,m_t)$.

Now let us turn to the more important cases, $\mut$ and $\muc$.  First
consider the variation of $\eta_3^\star$ with respect to $\mut$, which
is displayed in \fig{fig:e3-mutW}.
\begin{nfigure}
\includegraphics[clip,width=\miniwidth]{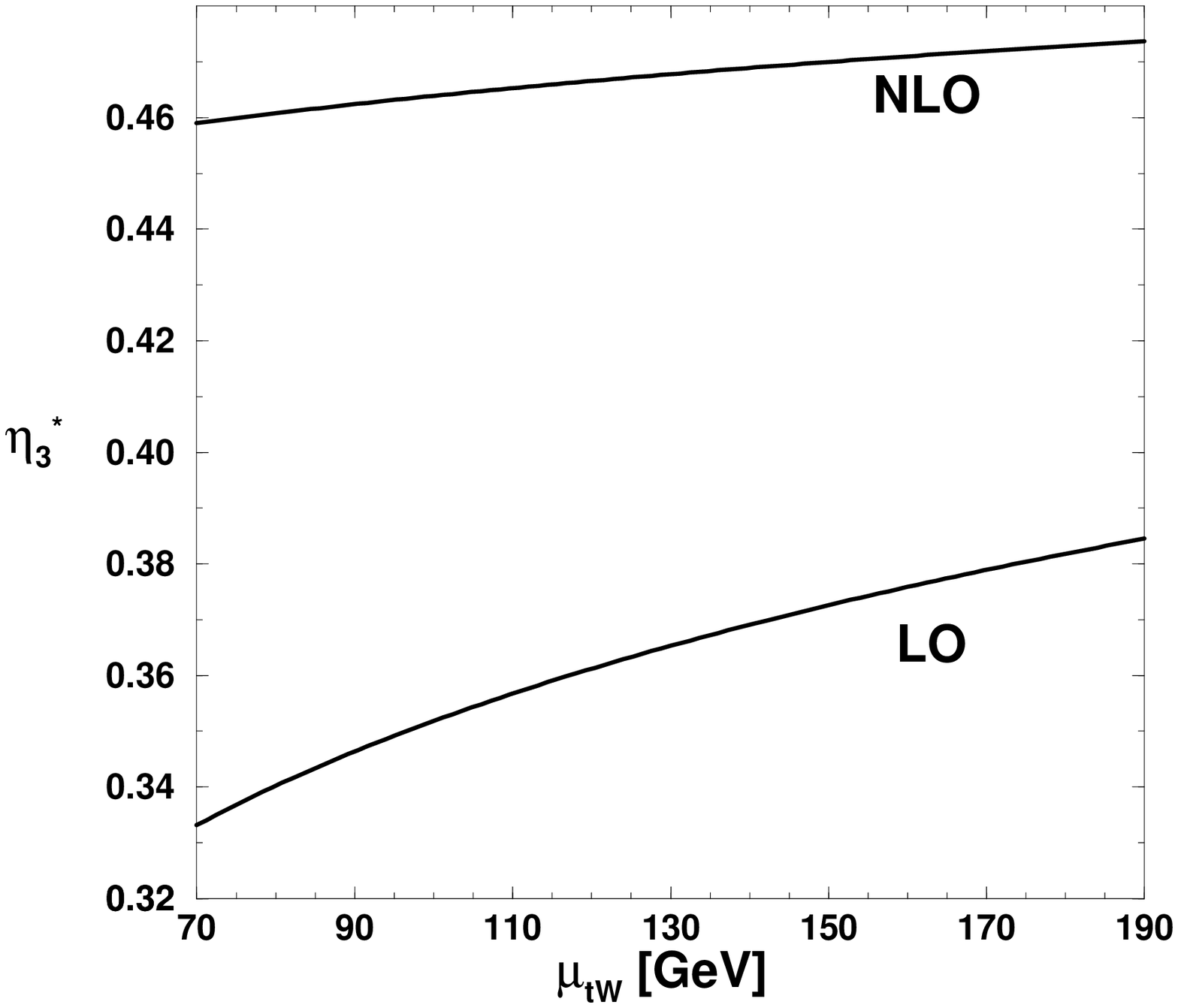}
\ncaption{The variation of $\eta_3^\star$ in LO and NLO with respect
to the scale $\mut$, at which the initial condition is defined.  The
other input parameters are given in \eq{InputParam}.}
\label{fig:e3-mutW}
\end{nfigure}
Since at $\mu_{tW}$ the top quark and the W-boson are integrated out
simultaneously, it is natural to choose the interval
$M_W\leq\mu_{tW}\leq m_t$ for the analysis.  In the LO result for
$\eta_3^\star$ we find a sizeable scale dependence of 12\%.  It is
almost totally removed in the NLO, where we obtain a variation of less
than 3\% in this interval.  This shows that it is very accurate to
integrate out the two heavy particles simultaneously. The strong
improvement in the NLO is due to the smallness of $\ln x_t$.

The situation is not so nice in the case of the variation of $\mu_c$,
which is displayed in \fig{fig:e3-muc}.
\begin{nfigure}
\includegraphics[clip,width=\miniwidth]{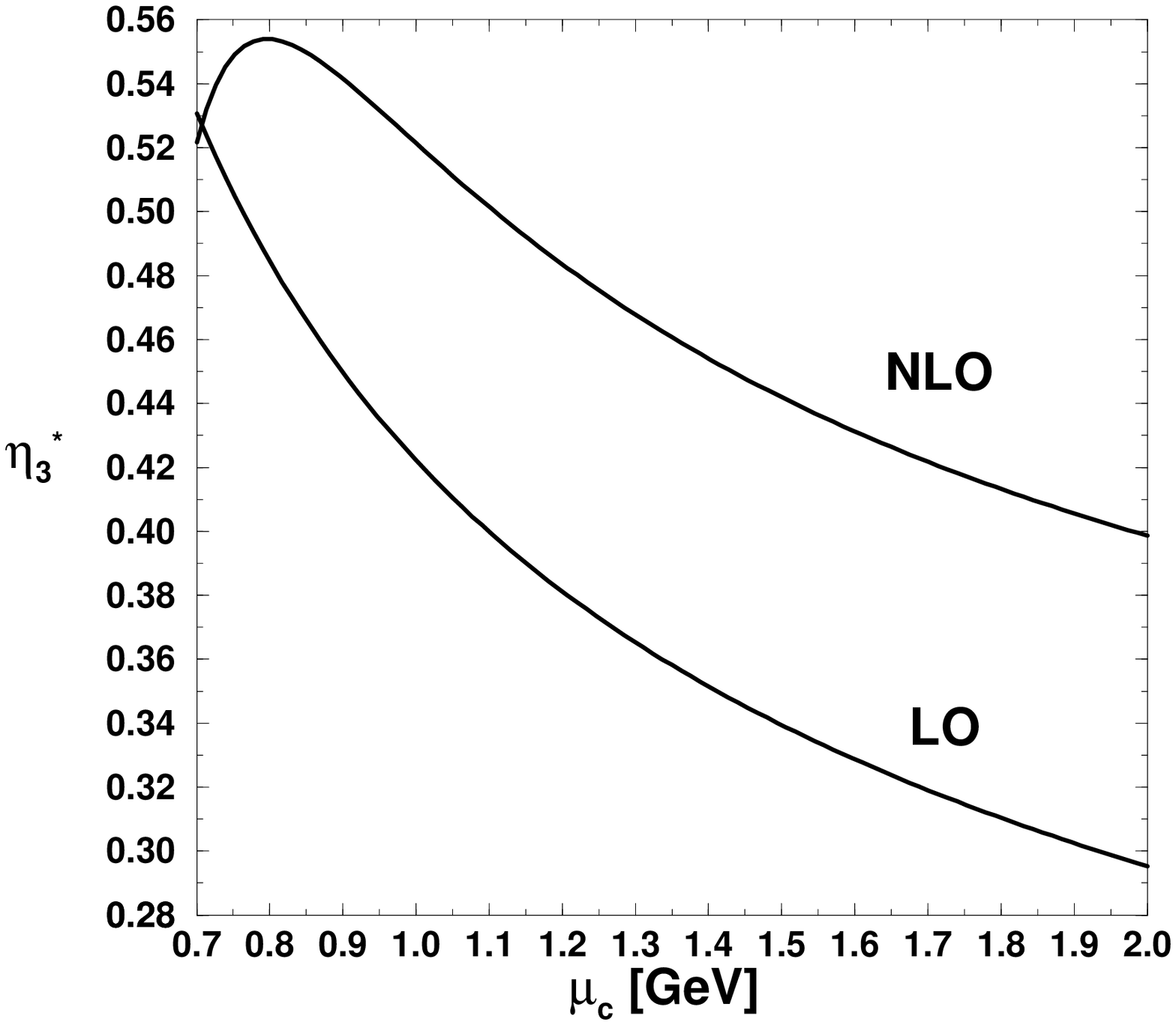}
\ncaption{The variation of $\eta_3^\star$ in LO and NLO with respect
to the scale $\muc$.  The range for the latter is taken unphysically
large to visualize the breakdown of perturbation theory.  The other
input parameters are given in \eq{InputParam}.}
\label{fig:e3-muc}
\end{nfigure}
We have intentionally extended
the range for $\mu_c$ to the unphysical low value of $0.7\gev$ to
visualize the breakdown of perturbation theory.  Varying $\mu_c$
within the interval $1.1\gev\leq\mu_c\leq1.6\gev$ yields
\begin{equation}
0.33 \leq {\eta_3^\star}^\mathrm{LO} \leq 0.40,
\hspace{1em}
0.43 \leq {\eta_3^\star}^\mathrm{NLO} \leq 0.50 .
\label{num-muc}
\end{equation}
This corresponds to a reduction of the scale dependence from 20\% to
14\%.  One reason for the poor improvement is the fact that the NLO
running of the mass is stronger than the LO one.

\subsection{Dependence of $\eta_3^\star$ on Physical Quantities}
\label{sect:num-phys}

Let us now investigate the dependence of $\eta_3^\star$ on the
physical parameters.  From the smallness of the coefficient $\cloc$ at
the initial scale one expects $\eta_3^\star$ to be almost independent
of $m_t^\star=m_t(m_t)$.  This statement is confirmed numerically,
allowing to treat $\eta_3^\star$ as $m_t$-independent in
phenomenological analyses.

The LO result for $\eta_3^\star$ depends on $m_c^\star=m_c(m_c)$
sizeably.  Yet this dependence is washed out nearly completely if one
looks at the NLO $\eta_3^\star$, see \fig{fig:e3-mc}.
\begin{nfigure}
\includegraphics[clip,width=\miniwidth]{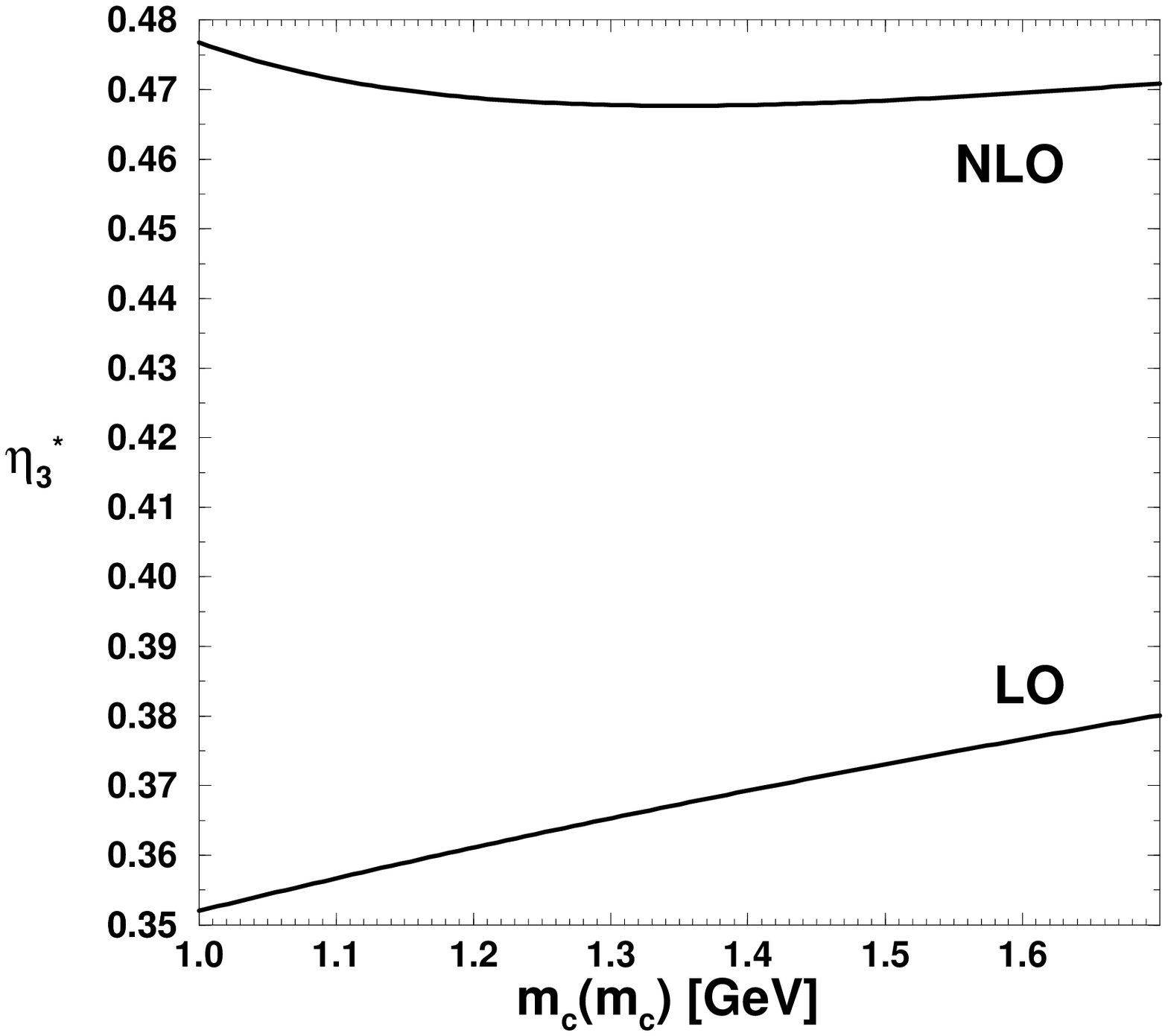}
\ncaption{The dependence of $\eta_3^\star$ on $m_c(m_c)$ in LO and
NLO.  The other input parameters are given in \eq{InputParam}.}
\label{fig:e3-mc}
\end{nfigure}

We close this section by a look at the dependence of $\eta_3^\star$ on
$\laQCD$, which is plotted in \fig{fig:e3-laQCD}.  It also turns out
to be very moderate.
\begin{nfigure}
\includegraphics[clip,width=\miniwidth]{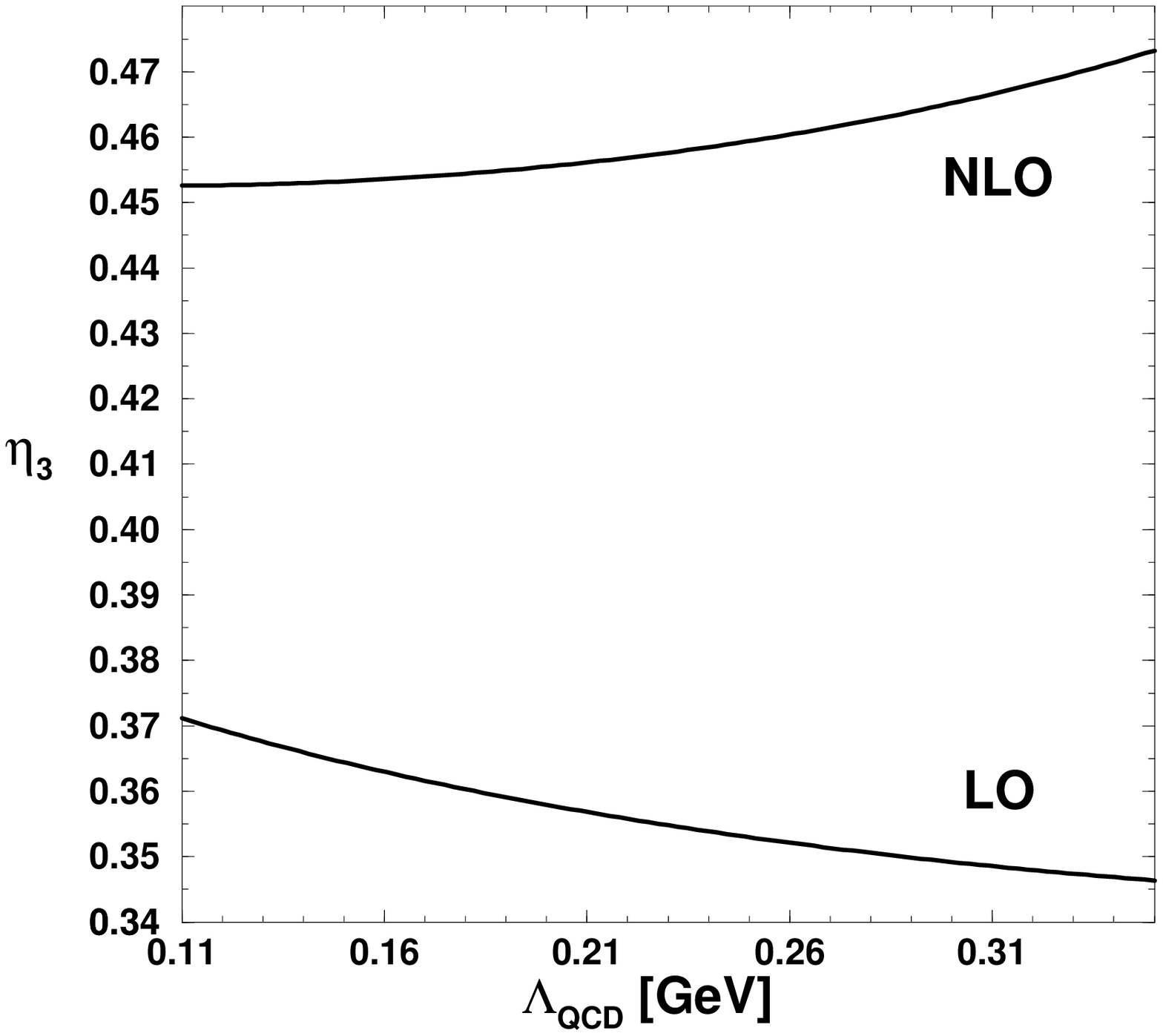}
\ncaption{The dependence of $\eta_3^\star$ on $\laQCD$ in LO and NLO.
Actual values for $\as (M_Z)$ correspond to $\laQCD^\mathrm{LO}
\approx 0.15\gev$ and $\laMSb \approx 0.31\gev$.  The other input
parameters are listed in \eq{InputParam}.}
\label{fig:e3-laQCD}
\end{nfigure}

\section{Conclusions}\label{sect:concl}

We have calculated the QCD short distance coefficient $\eta_3^\star$
of the low energy \dstwo\ hamiltonian in the next-to-leading order
(NLO) of renormalization group improved perturbation theory.  It
reads
\begin{equation}
\eta_3^{\star} = 0.47\errorpm{0.03}{0.04}\, ,
\quad
\eta_3^{\star \,\mathrm{LO} } \approx 0.37\, .
\label{coneta}
\end{equation}
The coefficient is scheme independent except that it depends on the
definition of the quark masses in $H^\dstwo$. The result in
{\eq{coneta}} corresponds to $\ov{\rm MS}$-masses $m_c(m_c)$ and
$m_t(m_t)$ as indicated by the superscript ``$\star$''.

The result has passed several checks:
\begin{enumerate}
\item
The NLO anomalous dimension tensor $\gbi{\pm j}$ {\eq{ResAnomTens1}}
has been found independent of the infrared structure of the two-loop
diagrams.
\item
We have kept the gluon gauge parameter $\xi$ arbitrary. It has
vanished from $\gbi{\pm j}$ after adding the contributions of the
diagrams with their correct combinatorial weight.
\item 
$\ln (m_c/\mu)/\eps$-terms have disappeared from the sum of two-loop
diagrams and counterterm diagrams.
\item 
The dependences of the final result for $\eta_3^\star$ on the matching
scales $\mu_{tW}$, $\mu_b$ and $\mu_c$ cancel to order $\as$.
Numerically the dependence has decreased.
\item 
If one expands the final result in powers of $\as$, one recovers the
terms proportional to $\as^0 \ln^1 x_c$, $\as^0 \ln^0 x_c$, $\as^1
\ln^2 x_c$, $\as^1 \ln^1 x_c$ of the result without RG improvement.
\item 
The initial condition for $\cloc$ in \eq{clocInit} as well as the
anomalous dimension tensor $\gbi{\pm j}$ in \eq{ResAnomTens1} depend
on the definition of the evanescent operators \eq{StdEvas}. We have
checked that this dependence is in accordance with the theorems of
\cite{hn2}, so that the final result is independent of the choice of
the evanescent operators.
\end{enumerate}



\begin{thebibliography}{10}
\bibitem{gw} F.\ J.\ Gilman and M.\ B.\ Wise, \prd 27 (1983) 1128.
\bibitem{bjw} A.\ J.\ Buras, M.\ Jamin and P.\ H.\ Weisz, \npb 347 (1990) 491.
\bibitem{hn1}  \hn, \npb 419 (1994) 292.
\bibitem{hn} \hn,
	\textsl{The Complete \dstwo-Hamiltonian in the Next-To-Leading
		Order}, \prp \textbf{hep-ph/9604330}, DESY 96-048,
		TUM-T31-86/96.
\bibitem{bw} A.~J.~Buras and P.~H.~Weisz, \npb 333(1990)66.
\bibitem{bjlw} A.~J.~Buras, M.~Jamin, M.~E.~Lautenbacher and P.~H.~Weisz,
	\npb 370 (1992) 69, addendum \npb 375 (1992) 501.
\bibitem{bjlw2} A.\ J.\ Buras, M.\ Jamin, M.\ E.\ Lautenbacher and 
	P.\ H.\ Weisz, \\ \npb 400 (1993) 37-74.
\bibitem{dg} M.~J.~Dugan and B.~Grinstein, \pl B256 (1991) 239.
\bibitem{hn2} \hn, \npb 455 (1995) 39-58.
\end{thebibliography}
\end{document}